\documentstyle[12pt]{article}
\catcode`@=11
\newcount\@tempcntc
\def\@citex[#1]#2{\if@filesw\immediate\write\@auxout{\string\citation{#2}}\fi
  \@tempcnta\z@\@tempcntb\m@ne\def\@citea{}\@cite{\@for\@citeb:=#2\do
    {\@ifundefined
       {b@\@citeb}{\@citeo\@tempcntb\m@ne\@citea\def\@citea{,}{\bf ?}\@warning
       {Citation `\@citeb' on page \thepage \space undefined}}%
    {\setbox\z@\hbox{\global\@tempcntc0\csname b@\@citeb\endcsname\relax}%
     \ifnum\@tempcntc=\z@ \@citeo\@tempcntb\m@ne
       \@citea\def\@citea{,}\hbox{\csname b@\@citeb\endcsname}%
     \else
      \advance\@tempcntb\@ne
      \ifnum\@tempcntb=\@tempcntc
      \else\advance\@tempcntb\m@ne\@citeo
      \@tempcnta\@tempcntc\@tempcntb\@tempcntc\fi\fi}}\@citeo}{#1}}
\def\@citeo{\ifnum\@tempcnta>\@tempcntb\else\@citea\def\@citea{,}%
  \ifnum\@tempcnta=\@tempcntb\the\@tempcnta\else
   {\advance\@tempcnta\@ne\ifnum\@tempcnta=\@tempcntb \else \def\@citea{--}\fi
    \advance\@tempcnta\m@ne\the\@tempcnta\@citea\the\@tempcntb}\fi\fi}
\catcode`@=12

\def\theequation{\arabic{section}.\arabic{equation}}
\def\barr{\begin{array}}
\def\earr{\end{array}}
\def\beq{\begin{equation}}
\def\eeq{\end{equation}}
\def\bea{\begin{eqnarray}}
\def\eea{\end{eqnarray}}
\def\bmath{\begin{displaymath}}
\def\emath{\end{displaymath}}
\def\bq{\begin{quote}}
\def\eq{\end{quote}}
\def\cA{{\cal A}}

\def\cL{{\cal L}}

\def\cP{{\cal P}}
\def\cT{{\cal T}}
\def\li{\lambda_i}
\def\lk{\lambda_k}
\def\ln{\lambda_n}
\def\lj{\lambda_j}
\def\lh{\lambda_h}
\def\ld{\lambda_\delta}
\def\PL{\mbox{P}_L}
\def\PR{\mbox{P}_R}
\def\apprle{\hspace{-0.1cm}\stackrel{\displaystyle <}{\sim}}
\def\apprge{\hspace{-0.1cm}\stackrel{\displaystyle >}{\sim}}
\def\slash#1{\setbox0=\hbox{$#1$}#1\hskip-\wd0\hbox to\wd0{\hss\sl/\/\hss}}
\def\snG{\mbox{{\footnotesize n}}_G}

\def\nG{\mbox{n}_G}

\def\veps{\varepsilon}
\def\g5{\gamma_5}
\def\<{\langle}
\def\>{\rangle}
\def\g{\mbox{g}}

\def\lZ{\lambda_Z}
\def\lS{\lambda_S}
\def\lR{\lambda_R}
\def\lI{\lambda_I}
\def\Li2{\mbox{Li$_2$}}

\def\lN1{\lambda_{N_1}}
\def\anp#1{{\em Ann.\ Phys.\ }{\bf #1}}

\def\mpla#1{{\em Mod.\ Phys.\ Lett.\ }{\bf A#1}}

\def\npb#1{{\em Nucl.\ Phys.\ }{\bf B#1}}
\def\plb#1{{\em Phys.\ Lett.\ }{\bf A#1}}
\def\plb#1{{\em Phys.\ Lett.\ }{\bf B#1}}
\def\prl#1{{\em Phys.\ Rev.\ Lett.\ }{\bf #1}}

\def\prd#1{{\em Phys.\ Rev.\ }{\bf D#1}}

\def\prep#1{{\em Phys.\ Rep.\ }{\bf #1}}
\def\ptp#1{{\em Prog.\ Theor.\ Phys.\ }{\bf #1}}

\def\zpc#1{{\em Z.\ Phys.\ }{\bf C#1}}
\begin{document}

\begin{flushright}
RAL/95-014\\
February 1995
\end{flushright}

\begin{center}
{\bf{\LARGE Confronting Left-Right Symmetric Models}}\\[0.4cm]
{\bf{\LARGE with Electroweak Precision Data}}\\[0.4cm]
{\bf{\LARGE at the {\em Z} Peak}}\\[2.cm]
{\large Apostolos Pilaftsis}\footnote[1]{E-mail address:
pilaftsis@v2.rl.ac.uk}\\[0.4cm]
{\em Rutherford Appleton Laboratory, Chilton, Didcot, Oxon, OX11 0QX, UK}
\end{center}
\vskip1.5cm
\centerline {\bf ABSTRACT}
\noindent
In view of  the  recent  and  future  electroweak precision
data accumulated at LEP and SLC,  we systematically analyze
possible new physics effects that may occur in the leptonic
sector within the context of $SU(2)_R \times SU(2)_L \times
U(1)_{B-L}$ theories. It is shown  that  nonobservation  of
flavour-violating  $Z$-boson  decays,  lepton  universality
in  the decays $Z\to l\bar{l}$, and universality of  lepton
asymmetries at the $Z$ peak form  a  set  of  complementary
observables, yielding severe  constraints on the  parameter
space of these theories. Contributions of new-physics effects
to $R_b=\Gamma(Z\to b\bar{b})/\Gamma(Z\to \mbox{hadrons})$
are found  to  give  interesting  mass  relations  for  the
flavour-changing Higgs scalars present in these models.

\newpage

\section{Introduction}
\indent

The Large Electron Positron collider (LEP) at CERN and the Stanford
Linear Collider (SLC) are powerful $e^+e^-$ machines operating
at the $Z$ peak, which can confront theoretical predictions of
the minimal Standard Model (SM) with experimental results to a
high accuracy.
A full analysis of all the electroweak precision data
including those of the year 1995 will either establish the SM up
to one-loop electroweak level or signal the onset of new physics.
In this context, analyzing electroweak oblique parameters~\cite{STU}
has become a common strategy to test the viability of models beyond
the SM. The electroweak oblique parameters are sensitive physical
quantities, when the new-physics interactions couple predominantly
to $W$ and $Z$ bosons. However, it is imperative to explore
additional observables that could be particularly sensitive to
other sectors of the SM.

In this paper, we will study a new set of {\em complementary }
leptonic observables and explicitly demonstrate the severe limitations
that can impose on model building of three-generation extensions of the SM.
The set of observables comprises flavour-changing leptonic decays
of the $Z$ boson~\cite{LFCNC,KPS}, universality-breaking parameters
$U_{br}$ for the diagonal decays $Z\to l\bar{l}$~\cite{BKPS}, and
universality-violating parameters $\Delta\cA_{l_1l_2}$ based on
lepton asymmetries measured at LEP and/or SLC~\cite{BP}.
For the sake of illustration, we will consider a minimal left-right
symmetric model (LRSM)~\cite{MS} described in Section~2.
Such a model can naturally generate vector--axial (V--A) as well
as V+A flavour-dependent $Zl\bar{l}$ couplings leading to
new physics effects that can be probed via the leptonic observables
mentioned above.
This set of observables will be discussed in some detail in Section 3.
In Section 4, we will give numerical estimates of these leptonic
observables within the framework of a minimal LRSM and investigate
their potential of effectively constraining this model. Furthermore,
attention will be paid to possible LRSM contributions to $R_b$.
Section 5 contains our conclusions.

\setcounter{equation}{0}
\section{The LRSM}
\indent

Left-right symmetric theories based on the gauge group $SU(2)_R\times
SU(2)_L\times U(1)_{B-L}$~\cite{PS,MS} were motivated from the fact that
the spontaneous breakdown of gauge and discrete symmetries can be
accomplished on the same footing. Such models can naturally arise
from $SO(10)$ grand unified theories via the breaking pattern~\cite{PS,FM}
\bmath
SO(10) \to SU(4)_{PS}\times SU(2)_R\times SU(2)_L \to SU(3)_c\times SU(2)_R
\times SU(2)_L \times U(1)_{B-L} \to \mbox{SM}.
\emath
We will, however, focus our analysis on the LRSM described in~\cite{MS}.

In the LRSM, right-handed neutrinos together with the
right-handed charged leptons form 3 additional weak isodoublets in
a three generation model. The classification of the quark sector
proceeds in an analogous way. To be specific, the assignment of quantum
numbers to fermions under the gauge group $SU_R(2)\times SU(2)_L \times
U(1)_{B-L}$ is arranged as follows:
\bea
L_L' = \left( \barr{c} \nu'_l \\ l' \earr \right)_L\ :\ (0,1/2,-1) &&\qquad
L_R' = \left( \barr{c} \nu'_l \\ l' \earr \right)_R\ :\ (1/2,0,-1),\\
Q_L' = \left( \barr{c} u' \\ d' \earr \right)_L\ :\ (0,1/2,1/3) &&\qquad
Q_R' = \left( \barr{c} u' \\ d' \earr \right)_R\ :\ (1/2,0,1/3),
\eea
where the prime superscript of the fermionic fields simply denotes
weak eigenstates.
In order to break the left-right gauge symmetry down to $U(1)_{em}$~\cite{MS},
we have to introduce one Higgs bidoublet,
\beq
\Phi\ =\ \left( \barr{cc} \phi^0_1 & \phi^+_1 \\
                          \phi^-_2 & \phi^0_2 \earr \right),
\eeq
that transforms as $(1/2^*,1/2,0)$ and two complex Higgs triplets,
\beq
\Delta_L\ =\ \left( \barr{cc} \delta_L^+/\sqrt{2} & \delta_L^{++} \\
                               \delta_L^0 & -\delta_L^+/\sqrt{2}
\earr \right)\qquad \mbox{and}\qquad
\Delta_R\ =\ \left( \barr{cc} \delta_R^+/\sqrt{2} & \delta_R^{++} \\
                               \delta_R^0 & -\delta_R^+/\sqrt{2}
\earr \right),
\eeq
with quantum numbers $(0,1,2)$ and $(1,0,2)$, respectively. For simplicity,
we will consider that only $\<\phi^0_1\>=\kappa_1/\sqrt{2}$ and
$\<\delta^0_R\>=v_R/\sqrt{2}$ acquire vacuum expectation values (vev's).
In practice, this can be accomplished by imposing invariance of the general
Higgs potential under judicious discrete symmetries of the bidoublet $\Phi$
and the Higgs triplets $\Delta_{L,R}$~\cite{IL}.
In fact, we will concentrate on case~(d) of Ref.~\cite{GGMKO},
to which the reader is referred for more details.
In case (d), it is $\<\delta^0_L\>= \<\phi^0_2\>=0$,
implying that the charged gauge bosons $W_L$ and $W_R$ represent
also physical states with masses $M_L=M_W$ and $M_R$, respectively. The
massive neutral gauge bosons $Z_L$ and $Z_R$ mix one another with a small
mixing angle of order $\kappa_1^2/v_R^2\sim 10^{-2}$.
To a good approximation, we will therefore neglect
$Z_L-Z_R$ mixing effects in our calculations.

Such a minimal LRSM allows the presence of baryon--lepton ($B-L$)
violating operators in the Yukawa sector. In fact, the
$B-L$-violating interactions are introduced by the triplet fields
$\Delta_{L,R}$ and give rise to Majorana mass terms $m_{M_{ij}}$
in the following
way:
\beq
\cL^{B-L}_{int} = - \frac{\sqrt{2} m_{M_{ij}}}{2v_R}\Big(h_{ij}
\bar{L}'^C_{L_i}\, \veps_{ij} \Delta_L L_{L_j}'\ +\
\bar{L}'^C_{R_i}\, \veps_{ij} \Delta_R L_{R_j}'\Big)\quad +\quad \mbox{H.c.}
\eeq
Here, $\veps_{ij}$ stands for the usual Levi-Civita tensor and the parameters
$h_{ij}=1$ if left-right symmetry is forced explicitly. However, a
phenomenological analysis of muon and $\tau$ decays shows that
$h_{ij}\ll 1$~\cite{GGMKO}. As a result,
$\delta_L^+$ and $\delta^{++}_L$ loop effects have been found to be
negligible~\cite{MAPS}.

In case (d), the neutrino mass matrix takes the general seesaw-type
form
\beq
M^\nu\ =\ \left( \barr{cc} 0 & m_D\\ m_D^T & m_M \earr \right),\label{Mnu}
\eeq
where $M^\nu$ is $6 \times 6$-dimensional matrix. In Eq.~(\ref{Mnu}), $m_D$
is a Dirac mass term connecting the left-handed neutrinos with the
right-handed ones. Relevant theoretical and phenomenological aspects
related to such neutrino mass models may be found in Ref.~\cite{ZPC}.
The matrix $M^\nu$ can always be diagonalized by a unitary $6\times 6$
matrix $U^\nu$ according to the common prescription $U^{\nu T}M^\nu U^\nu
=\hat{M}^\nu$. After diagonalization, one gets
6 physical Majorana neutrinos $n_i$ through the
transformations
\beq
\left(\barr{c} \nu_L' \\ \nu'^{C}_R \earr \right)_i\ =\
\sum_{j=1}^{2\snG} U^{\nu\ast}_{ij}\ n_{Lj}\quad \mbox{and}\quad
\left(\barr{c} \nu'^C_L \\ \nu'_R \earr \right)_i\ =\
\sum_{j=1}^{2\snG} U^\nu_{ij}\ n_{Rj}.
\eeq
The first $\nG =3$ neutral states, $\nu_i$ ($\equiv n_i$ for $i=1,\dots,\nG$),
are identified with the known $\nG$ light neutrinos,
while the remaining $\nG$ mass eigenstates, $N_j$ ($\equiv n_{j+\snG}$ for
$j=1,\dots,\nG$), are heavy Majorana neutrinos predicted by the model.
In addition to the leptonic sector, the quark sector of such an extension
contains non-SM couplings of the fermionic fields to the gauge and Higgs
bosons. Part of the LRSM couplings has been listed in~Ref.~\cite{IL,GGMKO,GZ}.
Relevant Feynman rules and additional discussion is given in Appendix A.

Adopting the conventions of Ref.~\cite{ZPC},
the interactions of the Majorana neutrinos, $n_i$, and charged
leptons, $l_i$, with the gauge bosons $W^\pm_L\ (\equiv W^\pm)$ and $Z_L$
are correspondingly mediated by the mixing matrices
\beq
B^L_{lj}\ = \sum\limits_{k=1}^{\snG} V^L_{lk} U^{\nu\ast}_{kj}\quad
\mbox{and}\quad
C^L_{ij}\ =\ \sum\limits_{k=1}^{\snG}\ U^\nu_{ki}U^{\nu\ast}_{kj}, \label{BL}
\eeq
with $l=1,\dots,\nG$ and $i,j=1,\dots,2\nG$. By analogy,
there exist mixing matrices $B^R_{li}$ and $C^R_{ij}$ given by
\beq
B^R_{lj}\ = \sum\limits_{k=\snG +1 }^{2\snG} V^R_{lk} U^{\nu\ast}_{kj}\quad
\mbox{and}\quad
C^R_{ij}\ =\ \sum\limits_{k=\snG +1 }^{2\snG}\ U^\nu_{ki}U^{\nu\ast}_{kj},
\label{BR}
\eeq
which are responsible for the couplings of $W^\pm_R$ and $Z_R$ to
charged leptons and Majorana neutrinos. In Eqs.~(\ref{BL}) and (\ref{BR}),
the unitary $\nG\times \nG$-matrices $V^L$ and $V^R$ are responsible
for the diagonalization of the charged lepton mass matrix via biunitary
transformations.

Due to the specific structure of $M^\nu$ in Eq.~(\ref{Mnu}),
the flavour-mixing matrices $B^L$ and $C^L$
satisfy a number of identities that may be found in~\cite{APetal}
These identities, which result from the requirement of unitarity
and renormalizability of the theory, turn out to be very useful
in deriving model-independent relations between the mixings $B^L_{li}$,
$C^L_{ij}$ and heavy neutrino masses. In a two generation mixing model,
we have~\cite{AP1,IP}
\beq
B^L_{lN_1}\ =\ \frac{\rho^{1/4} s^{\nu_l}_L}{\sqrt{1+\rho^{1/2}}}\ , \qquad
B^L_{lN_2}\ =\ \frac{i s^{\nu_l}_L}{\sqrt{1+\rho^{1/2}}}\ ,
\eeq
where $\rho=m^2_{N_2}/m^2_{N_1}\ (\ge 1)$
is a mass ratio of the two heavy Majorana
neutrinos $N_1$ and $N_2$ present in such a model,
and $s^{\nu_l}_L$ is $L$-violating mixings defined as~\cite{LL}
\beq
(s^{\nu_l}_L)^2 \ \equiv\  \sum\limits_{i=\snG +1}^{2\snG} |B^L_{li}|^2
\ \simeq\ \left( m_D^\dagger \frac{1}{m_M^2} m_D \right)_{ll}.
\eeq
Furthermore, the mixings $C^L_{N_iN_j}$ are determined by
\bea
C^L_{N_1N_1} &=& \frac{\rho^{1/2}}{1+\rho^{1/2}}\ \sum\limits_{i=1}^{\snG}
(s^{\nu_i}_L)^2, \qquad C^L_{N_2N_2}\ =\ \frac{1}{1+\rho^{1/2}}\
\sum\limits_{i=1}^{\snG} (s^{\nu_i}_L)^2, \nonumber\\
C^L_{N_1N_2}&=& -C^L_{N_2N_1}\ =\ \frac{i\rho^{1/4}}{1+\rho^{1/2}}\
\sum\limits_{i=1}^{\snG} (s^{\nu_i}_L)^2.
\eea

At this stage, it should be noted that $M^\nu$ of Eq.~(\ref{Mnu})
takes the known seesaw form~\cite{YAN} in case $m_M\gg m_D$.
Nevertheless, this mass-scale hierarchy can dramatically be relaxed in a
two-family seesaw-type model, which can radiatively induce light-neutrino
masses
in agreement with experimental upper bounds~\cite{ZPC}. The light-heavy
neutrino mixings
$s^{\nu_l}_L$ of such scenarios  can, in principle, scale as
$s^{\nu_l}_L \sim m_D/m_M$ rather than the known seesaw relation
$s^{\nu_l}_L \sim \sqrt{m_{\nu_l}/m_N}$. This implies that
high Dirac components are allowed to be present in $M^\nu$
and only the ratio $m_D/m_M$ ($\sim s^{\nu_l}_L$) gets limited by a global
analysis of low-energy data. Recently, such an analysis has been
performed in Ref.~\cite{Cliff}, in which the combined effect of
all possible effective operators of the charged- and neutral-current
interactions is considered. Although a careful analysis
can provide some model-dependent caveats, we will, however,
consider the following conservative upper bounds for the $L$-violating
mixings:
\beq
(s^{\nu_e}_L)^2,\  (s^{\nu_\mu}_L)^2 \ <\ 0.010,\qquad
(s^{\nu_\tau}_L)^2\ <\ 0.040,\qquad \mbox{and}
\qquad (s^{\nu_e}_L)^2 (s^{\nu_\mu}_L)^2\ < \ 1.\ 10^{-8}.\label{mix}
\eeq
For example, the last constraint in Eq.~(\ref{mix}) comes from
the nonobservation of decays of the type $\mu\to e\gamma,\ eee$,
or the absence of $\mu -e$ conversion events in nuclei.

In LRSMs, the mixing matrices $B^L$, $C^L$, $B^R$ and $C^R$ obey
the useful relations
\beq
\sum\limits_{i=1}^{2\snG} B^L_{l_1i}B^R_{l_2i} =  0,\qquad
\sum\limits_{l=1}^{\snG} B^{R\ast}_{li}B^R_{lj} =  C^R_{ij},\qquad
C^{L\ast}_{ij}\ +\ C^R_{ij}\ = \ \delta_{ij}. \label{idR}
\eeq
In a two-generation mixing model, Eq.~(\ref{idR}) together with
Eq.~(\ref{BR}) can be used to obtain the mixings
\bea
B^R_{l_1N_1} &=& \cos\theta_R\sqrt{1-C^L_{N_1N_1}},\qquad
B^R_{l_2N_1} \ =\ -\sin\theta_R\sqrt{1-C^L_{N_1N_1}}\nonumber\\
B^R_{l_1N_2} &=& -\cos\theta_R\frac{C^L_{N_1N_2}}{\sqrt{1-C^L_{N_1N_1}}}
\, -\, i\sin\theta_R\left( \frac{(1-C^L_{N_1N_1})(1-C^L_{N_2N_2})-
|C^L_{N_1N_2}|^2 }{1-C^L_{N_1N_1}} \right)^{1/2},
\nonumber\\
B^R_{l_2N_2} &=& \sin\theta_R\frac{C^L_{N_1N_2}}{\sqrt{1-C^L_{N_1N_1}}}
\, -\, i\cos\theta_R\left(\frac{(1-C^L_{N_1N_1})(1-C^L_{N_2N_2})-
|C^L_{N_1N_2}|^2 }{1-C^L_{N_1N_1}} \right)^{1/2}. \ \ \
\eea
Consequently, the leptonic sector of this two generation scenario
depends only on five free parameters;
the masses of the two heavy Majorana neutrinos, $m_{N_1}$ and $m_{N_2}$
[or equivalently $m_{N_1}$ and $\rho$], the mixing angles $(s^{\nu_i}_L)^2$,
which are, however, constrained by low-energy data, and an unconstrained
Cabbibo-type angle $\theta_R$.

\setcounter{equation}{0}
\section{SLC and LEP observables}
\indent

In this section, we will define more precisely the framework of our
calculations. In the limit of vanishing charged lepton masses, the
amplitude responsible for the decay $Z\to l_1\bar{l}_2$
can generally be parametrized as
\beq
\cT_l\ =\ \frac{ ig_w}{ 2c_w}\,
\veps^\mu_Z\, \bar{u}_{l_1}\gamma_\mu [g^{l_1l_2}_L\PL\
+\ g^{l_1l_2}_R\PR ] v_{l_2}, \label{Ampl}
\eeq
where $g_w$ is the usual electroweak coupling constant, $\veps_Z^\mu$ is
the $Z$-boson polarization vector, $u\ (v)$ is the Dirac spinor of
the charged lepton $l_1\ (l_2)$,
$\PL(\PR)=(1-(+)\gamma_5)/2$, and $c^2_w=1-s^2_w=M^2_W/M^2_Z$.
In Eq.~(\ref{Ampl}), we have defined
\beq
g^{l_1l_2}_{L,R}=g^l_{L,R}+\delta g^{l_1l_2}_{L,R},\qquad
g^l_L =\sqrt{\rho_l}(1-2\bar{s}^2_w),\qquad g^l_R=-2\sqrt{\rho_l}\bar{s}^2_w,
\eeq
where $\rho_l$, $\bar{s}_w$, $\delta g^l_{L,R}\ (\equiv\delta g_{L,R}^{ll})$
are obtained beyond the Born
approximation and are renormalization scheme dependent. It should be noted
that $\rho_l$, $\bar{s}_w$ introduce universal oblique corrections~\cite{STU},
whereas $\delta g^{l_1l_2}_{L,R}$ are flavour dependent. Obviously, an
analogous expression will be valid for the decay $Z\to b\bar{b}$, as soon as
$b$-quark mass effects can be neglected.

It is  convenient to reexpress the flavour-dependent electroweak
corrections in terms of the loop functions $\Gamma^L_{l_1l_2}$
and $\Gamma^R_{l_1l_2}$ as follows:
\bmath
\delta g^{l_1l_2}_L\ =\ \frac{\alpha_w}{2\pi} \, \Gamma^L_{l_1l_2}, \qquad
\delta g^{l_1l_2}_R\ =\ \frac{\alpha_w}{2\pi} \, \Gamma^R_{l_1l_2}.
\emath
The nonoblique loop functions $\Gamma^L_{l_1l_2}$
and $\Gamma^R_{l_1l_2}$ depend on whether the underlying theory
is of V--A or V+A nature. In Appendix B, we have analytically derived
the loop functions  $\Gamma^L_{l_1l_2}$ and $\Gamma^R_{l_1l_2}$ in the
context of LRSMs.
It is then straightforward to obtain the branching ratio for the possible
decay of the $Z$ boson into two different charged leptons
\beq
B(Z\to \bar{l}_1 l_2+l_1\bar{l}_2)\ =\ \frac{\alpha_w^3}{48\pi c^2_w}\,
\frac{M_Z}{\Gamma_Z} \Big[ |\Gamma^L_{l_1l_2}|^2+|\Gamma^R_{l_1l_2}|^2\Big],
\label{Blfv}
\eeq
with $\alpha_w=g^2_w/4\pi$.
Such an observable is constrained by LEP results to be, {\em e.g.},
$B(Z\to e\tau)\apprle 10^{-5}$~\cite{PDG}.

Another observable that has been introduced in~\cite{BKPS} is the
universality-breaking parameter $U_{br}^{(l_1l_2)}$.
To leading order of perturbation theory,
$U_{br}^{(l_1l_2)}$ is given by
\bea
U_{br}^{(l_1l_2)} &=& \frac{\Gamma (Z\to l_1\bar{l}_1)\ -\
\Gamma (Z\to l_2\bar{l}_2)}{\Gamma (Z\to l_1\bar{l}_1)\ +\
\Gamma (Z\to l_2\bar{l}_2)}\ - \ U_{br}^{(l_1l_2)}(\mbox{PS})\nonumber\\
&=& \frac{g^l_L (\delta g_L^{l_1} - \delta g^{l_2}_L)\ +\
g^l_R (\delta g^{l_1}_R - \delta g_R^{l_2} )}{g^{l2}_L\ +\ g^{l2}_R}\nonumber\\
&=& U_{br}^{(l_1l_2)}(\mbox{LH})\ +\ U_{br}^{(l_1l_2)}(\mbox{RH}), \label{Ubr}
\eea
where $U_{br}^{(l_1l_2)}(\mbox{PS})$ characterizes known phase-space
corrections coming from the nonzero masses of the charged leptons
$l_1$ and $l_2$ that can always be subtracted, and
\bea
U_{br}^{(l_1l_2)}(\mbox{LH})&=&
\frac{g^l_L(\delta g_L^{l_1}- \delta g_L^{l_2})}{(g^{l2}_L+g^{l2}_R)}\ =\
\frac{\alpha_w}{2\pi}\frac{g^l_L}{g^{l2}_L+g^{l2}_R}\, \Re e(\Gamma^L_{l_1l_1}
-\Gamma^L_{l_2l_2}) \\
U_{br}^{(l_1l_2)}(\mbox{RH})
&=&\frac{g^l_R(\delta g_R^{l_1}- \delta g_R^{l_2})}{g^{l2}_L+g^{l2}_R}\ =\
\frac{\alpha_w}{2\pi}\frac{g^l_R}{g^{l2}_L+g^{l2}_R}\, \Re e(\Gamma^R_{l_1l_1}
-\Gamma^R_{l_2l_2})
\eea
To make contact with the corresponding observable given in~\cite{PDG},
one can easily derive the relation
\bmath
\frac{\Gamma (Z\to l\bar{l})}{\Gamma (Z\to l'\bar{l}')}
= 2 U^{(ll')}_{br} + 1.
\emath

The results of a combined analysis at LEP/SLC regarding lepton universality
at the $Z$ resonance can be summarized~\cite{LEP,SLC}
as follows:
\bea
|U_{br}^{(ll')}| & < & 5.\ 10^{-3}\qquad (\mbox{SM}:0),\nonumber\\
\cA_\tau (\cP_\tau) &=& 0.143 \pm 0.010\qquad (\mbox{SM}:0.143),\nonumber\\
\cA_e (\cP_\tau) &=& 0.135 \pm 0.011,\nonumber\\
\cA_{FB}^{(0,l)} &=& 0.0170\pm 0.0016\qquad (\mbox{SM}:0.0153),\nonumber\\
\cA_{LR} (SLC)& =& 0.1637 \pm 0.0075.\label{Exp}
\eea
In parentheses, we quote the theoretical predictions obtained in the SM.
{}From (\ref{Exp}), we find that the experimental sensitivity to
$\Delta\cA_{\tau e}$ is about $7\%$, ($4\%$) for LEP (SLC).
Note that $\cA_e$ should equal the left-right asymmetry, $\cA_{LR}$,
measured at SLC.
Furthermore, it is worth mentioning that ongoing SLC experiments
are measuring the observable
\beq
A_{LR}^{FB}(f)\ =\ \frac{\Delta\sigma( e^-_Le^+\to f\bar{f})_{FB}
- \Delta\sigma( e^-_Re^+\to f\bar{f})_{FB}}{\Delta\sigma( e^-_Le^+\to
f\bar{f})_{FB} + \Delta\sigma( e^-_Re^+\to f\bar{f})_{FB}}=\frac{3}{4}
\cP_e \cA_f, \label{AFBLR}
\eeq
The forward-backward left-right asymmetry for individual flavours will
be an interesting alternative of establishing possible
deviations from SM universality in lepton asymmetries.

Lepton asymmetries [or equivalently forward-backward asymmetries]
can also play a crucial r\^ole to constrain new physics. Here, we will
be interested in experiments at LEP/SLC that measure the observable
\bea
\cA_l \ &=&\ \frac{\Gamma (Z\to l_L \bar{l})\ -\ \Gamma (Z\to l_R \bar{l})}{
\Gamma (Z\to l \bar{l})}\ =\ \frac{g^{ll2}_L\ -\ g^{ll2}_R}{g^{ll2}_L\ +\
g^{ll2}_R} \nonumber\\
 &=& \frac{ g^{l2}_L -g^{l2}_R + 2(g^l_L\delta g^l_L - g_R\delta g^l_R) }{
g^{l2}_L + g^{l2}_R + 2(g^l_L\delta g^l_L + g^l_R\delta g^l_R)}.\label{Al}
\eea
In view of the recent discrepancy of more than $2\sigma$
between the experimental results of $\cA_{LR}$ at SLC and
$\cA_e$ at LEP, we are motivated to use the nonuniversality
parameter of lepton asymmetries~\cite{BP}
\beq
\Delta\cA_{l_1l_2}\ =\ \frac{\cA_{l_1}\ -\ \cA_{l_2}}{\cA_{l_1}\ +\ \cA_{l_2}}
\ =\ \frac{1}{\bar{\cA}_l} \Big( U_{br}^{(l_1l_2)}(\mbox{LH})\ -\
U_{br}^{(l_1l_2)}(\mbox{RH}) \Big)\ -\ U_{br}^{(l_1l_2)}, \label{DA}
\eeq
where $\bar{\cA}_l$ may be given by the mean value of the
two lepton asymmetries $\cA_{l_1}$ and $\cA_{l_2}$.
At this point, it should be stressed that requiring $U_{br}^{(l_1l_2)}=0$ does
{\em not necessarily} imply $\Delta\cA_{l_1l_2}=0$. As we will later see,
in LRSMs one can naturally encounter the possibility, in which
$U_{br}(\mbox{LH})\simeq -U_{br}(\mbox{RH})$ while $\Delta\cA_{l_1l_2}$
becomes sizeable. Moreover, the physical quantities $U_{br}^{(l_1l_2)}$
and $\Delta\cA_{l_1l_2}$ do not depend explicitly on universal electroweak
oblique parameters, especially when the latter ones may poorly constrain
such three-generation scenarios~\cite{BKPS}.

Another observable which will still be of interest is
\beq
R_b \ =\ 0.2202 \pm 0.0020\qquad (\mbox{SM}:0.2158).
\eeq
If the measurement at LEP is correct, $R_b$ turns out to be about $2\sigma$
off from the theoretical prediction of the minimal SM.
New physics contributions to $R_b$ can be conveniently calculated
through~\cite{BDV}
\beq
R_b\ =\ 0.22\Big[ 1+0.78\nabla_b^{(SM)}(m_t)-0.06\Delta\rho^{(SM)}(m_t)\Big],
\eeq
where $\nabla_b^{(SM)}(m_t)$ and $\Delta\rho^{(SM)}(m_t)$ contains the
$m_t$-dependent parts of the vertex and oblique corrections, respectively.
Practically, only $\nabla_b^{(SM)}(m_t)$ gives significant negative
contributions to $R_b$, which behave, in the large top-mass limit,
as~\cite{BPS}
\beq
\nabla_b^{(SM)}(m_t)\ \simeq\ -\frac{20\alpha_w s^2_w}{13\pi}\,
\frac{m^2_t}{M^2_Z}.
\eeq
If there are new physics effects contributing to $\nabla_b^{(SM)}(m_t)$,
these can be estimated by
\beq
\nabla_b^{(new)}(m_t)\ =\ \frac{\alpha_w}{2\pi}\,
\frac{g^b_L\Re e(\Gamma^L_{bb}(m_t)-\Gamma^L_{bb}(0))
+g^b_R\Re e(\Gamma^R_{bb}(m_t)-\Gamma^R_{bb}(0))}{g^{b2}_L\ +\ g^{b2}_R}\, ,
\label{Rbnew}
\eeq
where $g^b_L=1-2s_w^2/3$ and $g^b_R=-2s^2_w/3$.
In the next section, we will analyze numerically the size of new physics
effects expected in LRSM.

\setcounter{equation}{0}
\section{Numerical results and discussion}
\indent

Since there is a large number of free parameters that could vary
independently in the LRSM, we have fixed to typical values
all of them except of one each time and investigated the behaviour
of our observables as a function of the remaining kinematic variable.
More explicitly, we have found that $\delta^{++}_R$ quantum corrections
to the effective $Zl_R\bar{l}_R$ coupling shown in Fig.~1 are very small,
since $M_{\delta^{++}_R}> 5$~TeV for phenomenological reasons~\cite{GGMKO}.
The very same lower mass bound should obey the flavour-changing scalars
$\phi^{0r}_2\ (\equiv\Re e(\phi^0_2)/\sqrt{2})$ and $\phi^{0i}_2\
(\equiv\Im m(\phi^0_2)/\sqrt{2})$~\cite{GGMKO}.
However, the mass difference between the two flavour-changing scalars
should not be too large because the latter would lead to large negative
contributions to $R_b$ (we will discuss the consequences from a large
mass-difference realization between the flavour-changing scalars
at the end of this section). In our estimates, we have
assumed that $\phi^{0r}_2$ and $\phi^{0i}_2$ are nearly degenerate
and heavier than 5 TeV. In such a case, loop effects involving
flavour-changing scalars are found to be vanishingly small.

In order to increase the predictability of our LRSM but still keep our
analysis on a general basis, we shall consider a two-generation
mixing scenario.
Then, the free parameters of our minimal model are: the lepton-violating
mixings $(s^{\nu_l}_L)^2$ [which are, however, constrained to some extend by
a global analysis of low-energy data], the two heavy neutrino masses
$m_{N_1}$ and $m_{N_2}$ [which have been taken to be at the same
mass scale $m_N$], the masses of the charged gauge boson $M_R$ and
its orthogonal associate scalar $M_h$, and a Cabbibo-type
angle $\theta_R$ that rotates the right-handed charged leptons to
the corresponding mass eigenstates.

In Figs.~2(a)--(d), we present plots of $B(Z\to e^-\tau^++e^+\tau^-)$
as a function of $m_N$, $M_R$, $M_h$, and $\theta_R$ while keeping fixed the
remaining kinematic parameters. In Fig.~2(a), we see the characteristic
{\em quadric}, $m^4_N/M^4_W$, dependence on the branching
ratio~\cite{KPS,APetal}.
The dashed, dotted and dash-dotted lines represent results coming purely
from the $SU(2)_R$ sector for $(s^{\nu_\tau}_L)^2=0.040,\ 0.030$, and
$0.020$, respectively. The solid lines~$i$, $ii$, and $iii$ correspond
to a complete computation for the three different lepton-violating
mixings mentioned above.
If we assume some typical values for
the rest of the parameters, {\em i.e.}~$M_R=0.4$~TeV, $M_h=30$~TeV,
and $\theta_R=0$, we find that $B(Z\to e^-\tau^+ + e^+\tau^-)\apprle
2.\ 10^{-6}$ for $m_N=3$~TeV. Although the size of new physics effects
may be probed at LEP, the reported value is
$B(Z\to e\tau)<10^{-5}$ and it does not yet impose rather severe constraints
on the parameter space of the theory. This conclusion is also supported
by Figs.~2(b)--(d). In Fig.~2(c), it is worth observing the logarithmic
dependence of the mass ratio $M_h/M_R$ on the branching ratio, which
can also render the decay channel $Z\to e\tau$ measurable. In Fig.~2(d),
one can further see the strong dependence of $\theta_R$ on
$B(Z\to e^-\tau^+ + e^+\tau^-)$. However, a similar, though complementary,
behaviour will be found to be present in the observables $U_{br}$ and
$\Delta\cA$.

We are now proceed by examining numerically the dependence of the
universality-breaking parameter $U_{br}$ as a function of various
kinematic variables shown in Figs.~3(a)--(d).
Again, we observe the nondecoupling behaviour of the heavy neutrino
mass in the observable $U_{br}$~\cite{BKPS}.
The size of new physics becomes significant for $m_N\apprge 3$ TeV,
{\em i.e.} $U_{br}\sim 4.-5.\ 10^{-3}$. In Fig.~3(b), we see that the
value of $U_{br}$ decreases rapidly as $M_R$ increases. In Fig.~3(c),
we remark again the logarithmic dependence $M^2_h/M^2_R$ on $U_{br}$.
In our estimates, we have used a Cabbibo-type angle $\theta_R=45^0$, which
turns out to be a rather moderate value as is displayed in Fig.~3(d).
As has been mentioned above, the electroweak corrections originating
genuinely from the $SU(2)_R$ sector depend on the angle $\theta_R$.
Looking at Fig.~3(d), one can readily see that the choice $\theta_R=45^0$
gives smaller effects of nonuniversality in the leptonic
partial widths of the $Z$ boson.
If we had chosen $\theta_R=-45^0$, we would have obtained much
stronger combined bounds on the mass parameters and $L$-violating
mixings of the LRSM.

One may get the impression that new-physics effects can be
minimized by selecting $\theta_R$ to lie in a specific range.
This is, however, not true, since the universality-breaking
parameter $\Delta\cA_{l_1l_2}$ will play a complementary r\^ole
as is shown in Figs.~4(a)--(d). In Fig.~4, we list the results
after adding both contributions coming from $SU(2)_L$ and $SU(2)_R$
gauge sectors.
Thus, we may be sensitive up to $m_N\apprle 1.5$~TeV for
$(s^{\nu_\tau}_L)^2=0.04$ and $(s^{\nu_e}_L)^2=0.01$ (see Fig.~4(a)).
In Fig.~4(b), we display the decoupling effect of a very heavy $W_R$.
In Fig.~4(c), we find again the logarithmic enhancement caused by
the nondegeneracy between $W_R^\pm$ and $h^\pm$.
In addition, it should be noted that interesting phenomenology can
only arise for relatively light $W_R$ bosons,
{\em i.e.}~$M_R\apprle 1$~TeV. The latter observation can also be verified
from Fig.~4(d), in which $\Delta\cA$ is drawn as a function of $\theta_R$
for $M_R=0.4,\ 0.6,$ and 0.8 TeV. Furthermore, one can easily recognize the
complementary r\^ole that $B(Z\to l_1l_2)$, $U_{br}$, and $\Delta\cA$
play as far as $\theta_R$ is concerned, when comparing Figs.~2(d), 3(d), and
4(d). For example, the choice $\theta_R=-45^0$ would make
$\Delta\cA$ more difficult to observe, whereas $U_{br}$ becomes larger
for this value of $\theta_R$. Of course,
scenarios where $M_R$ is at the TeV scale may not be
compatible with $K_L-K_S$ phenomenology if we assume an exact
left-right symmetry in the Yukawa sector of the model. Nevertheless,
in LRSMs that possess nonmanifest or pseudomanifest left-right
symmetry, such a constraint is not valid any longer~\cite{LSS}.

In the following, we will try to address the question whether
there exist possibilities of inducing positive contributions
to $R_b$ within our LRSM. As has already been noticed
in Section 3, only positive contributions to $R_b$ are of
potential interest, which will help to achieve a better agreement between
theoretical prediction and the experimental value of $R_b$. In LRSM,
we first consider the Feynman graphs 1(m) and 1(n), where the external
leptons are replaced by $b$-quarks and virtual down-type quarks are running
in the place of charged leptons. The interaction
Lagrangians of the flavour-changing scalars $\phi^{0r}_2$ and $\phi^{0i}_2$
with the $d,\ s,\ b$ quarks can be obtained by Eq.~(A.17) after making
the obvious replacements. These couplings are enhanced, as they are
proportional to the top-quark mass. In fact, the flavour-changing scalars
generate effective $Zb\bar{b}$ couplings of both V--A and V+A nature.
In the limit $M_{\phi^{0r}_2},\ M_{\phi^{0i}_2}\gg M_Z$,
the effective $Zb\bar{b}$ couplings take the simple form
\bea
\Re e(\Gamma^R_{bb}(m_t)-\Gamma^R_{bb}(0)) &=&  \frac{1}{8} |V^R_{tb}|^2
\frac{m^2_t}{M^2_W}\, \left( \,
\frac{\lambda_S +\lambda_I}{2(\lambda_S-\lambda_I)}\,
\mbox{ln}\frac{\lambda_S}{\lambda_I}\, -\, 1\, \right), \label{RbL}\\
\Re e(\Gamma^L_{bb}(m_t)-\Gamma^L_{bb}(0)) &=& - \frac{1}{8} |V^L_{tb}|^2
\frac{m^2_t}{M^2_W}\, \left( \, \frac{\lambda_S +\lambda_I}{2(\lambda_S
-\lambda_I)}\, \mbox{ln}\frac{\lambda_S}{\lambda_I}\, -\, 1\, \right),
\label{RbR}
\eea
where $\lambda_S$ and $\lambda_I$ are defined in Appendix B after Eq.~(B.1).
The analytic function in the parentheses of the r.h.s.\ of
Eqs.~(\ref{RbL}) and (\ref{RbR}) is always positive and equals zero
when the two scalars $\phi^{0r}_2,\ \phi^{0i}_2$ are degenerate.
Substituting Eqs.~(\ref{RbL}) and (\ref{RbR}) into Eq.~(\ref{Rbnew}),
one easily finds that the SM value of $R_b$ is further decreased.
This leads automatically to the restriction
\beq
M_{\phi^{0r}_2}\ \simeq\  M_{\phi^{0i}_2}. \label{phi}
\eeq
The mass relation~(\ref{phi}) has been used throughout our numerical
estimates.

Another place that may lead to positive contributions to $R_b$ are
due to diagrams similar to Figs.~1(h) and 1(d). Indeed, an analogous
calculation gives
\beq
\Re e(\Gamma^R_{bb}(m_t)-\Gamma^R_{bb}(0)) \ =\ -\frac{1}{4} |V^R_{tb}|^2
\frac{m^2_t}{M^2_W}\, s^2_\beta c^2_\beta\,
\left(\, \frac{\lambda_h +\lambda_R}{2(\lambda_h
-\lambda_R)}\, \mbox{ln}\frac{\lambda_h}{\lambda_R}\, -\, 1\, \right).
\label{Rbh+}
\eeq
However, the l.h.s.\ of Eq.~(\ref{Rbh+}) is proportional to $s^2_\beta=
M^2_W/M^2_R$ yielding a rather small effect. If we insist in cancelling
the negative SM vertex correction $\nabla_b^{SM}(m_t)$
through the contribution (4.4), we find the highly unnatural mass ratio
\bmath
\frac{M_h}{M_R}\ \simeq \ e^{70}.
\emath
The latter also demonstrates the difficulty of obtaining radiatively
positive contributions to $R_b$ within the LRSM. In Ref.~\cite{Rizzo},
it has been shown that $Z_L-Z_R$ mixing effects could help in producing
contributions of either sign to $R_b$. We will not pursue this
topic here.

\section{Conclusions}
\indent

We have shown that lepton-flavour violating $Z$-boson decays,
lepton universality in the decays $Z\to l\bar{l}$, and universality
of lepton asymmetries at LEP/SLC represent a set of complementary
observables and can hence impose severe limitations
on model-building in the leptonic sector. For our illustrative purposes,
we have considered a LRSM with two-generation mixing. We have found that the
observables $B(Z\to l_1l_2)$, $U_{br}^{l_1l_2}$, and $\Delta\cA_{l_1l_2}$
are sensitive to different parameter-space regions of this minimal scenario.
For instance, if $(s^{\nu_\tau}_L)^2=0.03$, $(s^{\nu_e}_L)^2=0.01$,
$M_R=0.4$ TeV, and $M_h=30$ TeV, then heavy neutrinos are found to
have masses that do not exceed 2 TeV for any value of the Cabbibo-type
angle $\theta_R$. On the other hand, constraints on new physics from $R_b$
prefer scenarios, in which flavour-changing scalars are degenerate
in mass.

It may be worth remarking again the fact that LRSMs
can naturally predict $U_{br}\simeq 0$, for some choice of parameters,
which could naively be interpreted that lepton universality is preserved
in nature. As has been shown in this paper, universality violation
can manifest itself in lepton asymmetries $\Delta\cA_{l_1l_2}$ as well.
This is, however, not an accidental feature of the LRSM but may have
a general applicability to unified models, such as supersymmetric
extensions of the SM~\cite{BP}. In general,
such theories can naturally generate both nonuniversal V--A and V+A
$Zf\bar{f}$-couplings yielding effects that may be detected by
current experiments at LEP and SLC.

\vskip2cm
\noindent
{\bf Acknowledgements.} I wish to thank Richardo Barbieri, Jose Bernab\'eu,
Roger Phillips, and John Thompson for discussions and comments.

\newpage
\setcounter{section}{0}
\def\theequation{\Alph{section}.\arabic{equation}}
\begin{appendix}
\setcounter{equation}{0}
\section{Feynman rules in the LRSM}
\indent
Although some of the Feynman rules required in our problem were given in
Ref.~\cite{IL,GGMKO}, we list all the relevant Feynman rules and Lagrangians
governing the interactions of the gauge and Higgs bosons with leptons and
neutrinos, as well as the trilinear couplings of the bosons. The covariant
derivative acts on the Higgs multiplets as follows:
\bea
D_\mu\Phi &=& \partial_\mu\Phi + i\frac{g_L}{2} \vec{\sigma}\vec{W}_{L\mu}\Phi
- i\frac{g_R}{2} \Phi \vec{\sigma} \vec{W}_{R\mu},\\
D_\mu\Delta_{R,L} &=& \partial_\mu\Delta_{R,L}  + i\frac{g_{R,L}}{2}
[\vec{\sigma} \vec{W}_{R,L\mu},\Delta_{R,L}]
+ig' B_\mu\Delta_{R,L},
\eea
where $\sigma_i$ are the known $2\times 2$ Pauli matrices, and $g_L$ ($g_R$)
are the $SU(2)_L$ ($SU(2)_R$) weak coupling constants which will be
set equal to $g_w=g_L=g_R$ ($g_w$ is the usual $SU(2)_L$ weak coupling constant
in the SM).
To facilitate our computational task, we will further assume that the
corresponding neutral gauge boson $Z_L$ is the $Z$  of the SM
to a good approximation.
Also, we will list the novel LRSM interactions together with the SM
couplings in order to avoid possible ambiguities between relative signs.

The trilinear couplings of gauge, Higgs, and would-be Goldstone bosons
may therefore be obtained by (all momenta flow into the vertex)
\bea
Z_{L\nu}(r)W_{L\lambda}^+(p)W_{L\mu}^-(q)&:&-ig_wc_w
f_{\mu\nu\lambda}(r,q,p),\\
Z_{L\nu}(r)W_{R\lambda}^+(p)W_{R\mu}^-(q)&:&
                             ig_w \frac{s^2_w}{c_w}f_{\mu\nu\lambda}(r,q,p),\\
Z_{L\nu}(r)G_L^+(p)W_{L\mu}^-(q)&:& -ig_w M_W\frac{s^2_w}{c_w}\g^{\mu\nu}, \\
Z_{L\nu}(r)G_R^+(p)W_{R\mu}^-(q)&:& ig_w M_W\frac{s^2_\beta-s^2_w}{s_\beta c_w}
                                                               \g^{\mu\nu}, \\
Z_{L\nu}(r)h^+(p)W_{R\mu}^-(q)&:& -ig_w M_W\frac{c_\beta}{c_w}\g^{\mu\nu}, \\
Z_{L\mu}(r)G^+_L(p)G_L^-(q)&:& -i\frac{g_w}{2c_w} (1-2s^2_w) (p-q)_\mu, \\
Z_{L\mu}(r)G^+_R(p)G_R^-(q)&:& -i\frac{g_w}{2c_w} (s^2_\beta-2s^2_w)
                                                                  (p-q)_\mu, \\
Z_{L\mu}(r)h^+(p)h^-(q)&:& -i\frac{g_w}{2c_w} (c^2_\beta-2s^2_w)
                                                                  (p-q)_\mu, \\
Z_{L\mu}(r)G^\pm_R(p)h^\mp(q)&:& \mp i\frac{g_w}{2c_w} s_\beta c_\beta
                                                                 (p-q)_\mu, \\
Z_{L\mu}(r)\phi^{0r}_2(p)\phi^{0i}_2(q)&:& \frac{g_w}{2c_w} (p-q)_\mu, \\
Z_{L\mu}(r)\delta^{++}_R(p)\delta^{--}_R(q)&:& 2i\frac{g_w}{c_w}s^2_w(p-q)_\mu.
\eea
Here, we have defined $s_\beta = \sqrt{1-c^2_\beta}=M_W/M_R$ and
the Lorentz tensor $f_{\mu\nu\lambda}(r,q,p)=(r-q)_\lambda\g_{\mu\nu} +
(q-p)_\nu \g_{\lambda\mu} + (p-r)_\mu\g_{\nu\lambda}$.

The corresponding couplings of the gauge, Higgs, and would-be Goldstone
bosons to the charged leptons and neutrinos can be read off from the
Lagrangians
\bea
\cL^{W_R}_{int} &=& -\frac{g_w}{\sqrt{2}} W_R^{-\mu}\, B^R_{li}\
\bar{l}\gamma_\mu \PR n_i\quad +\quad \mbox{H.c.},\\
\cL^{G_R^-}_{int} &=& -\frac{g_w}{\sqrt{2}M_W}\, s_\beta \, G_R^{-}\, B^R_{li}\
\bar{l}\Big[ m_l\PR - m_{n_i}\PL\Big] n_i\quad +\quad \mbox{H.c.},\\
\cL^{h^-}_{int} &=& \frac{g_w}{\sqrt{2}M_W}\, c_\beta\, h^{-}\,
\bar{l}\left[ B^R_{li} m_l\PR - B^R_{lj}\left(\delta_{ji}-
\frac{C^{R\ast}_{ji}}{c^2_\beta} \right)m_{n_j}\PL\right]n_i\quad
+\quad \mbox{H.c.},\ \\
\cL^{\phi_2^0}_{int} &=& -\frac{g_w}{2M_W} \phi_2^{0r}\,
\bar{l}_1\Big[ B^L_{l_1j}m_{n_j}B^{R\ast}_{l_2j}\PR +
B^R_{l_1j}m_{n_j}B^{L\ast}_{l_2j}\PL\Big]l_2\nonumber\\
&&-\frac{ig_w}{2M_W} \phi_2^{0i}\,
\bar{l}_1\Big[ B^L_{l_1j}m_{n_j}B^{R\ast}_{l_2j}\PR -
B^R_{l_1j}m_{n_j}B^{L\ast}_{l_2j}\PL\Big]l_2,\\
\cL^{\delta_R^{++}}_{int} &=& \frac{g_w}{2\sqrt{2}M_W} \frac{s_\beta}{c_\beta}
\delta_R^{++}\,
\bar{l}_1^C\, B^{R\ast}_{l_1j}m_{n_j}B^{R\ast}_{l_2j}\PR\, l_2 \quad
+\quad \mbox{H.c.},
\eea
where the mixing matrices $B^R$ and $C^R$ are defined in Section 2.
The couplings of $Z_L\ (\equiv Z)$ to Majorana neutrinos
may be found in Ref.~\cite{ZPC}.

\setcounter{equation}{0}
\section{The nonoblique {\boldmath $Zl\bar{l}$} vertex}
\indent

We analytically evaluate the loop amplitudes in the limit of vanishing
external lepton masses. We adopt dimensional regularization in conjunction
with the reduction algorithm of Ref.~\cite{PV}. Unlike the metric notation
of Ref.~\cite{PV}, we use the Minskowskian metric,
$\g^{\mu\nu}=\mbox{diag}(1,1,...,-1)$.

The nonoblique effective $Zll'$ vertex function is similar to the one
obtained in~\cite{KPS}. Its analytic form is given by (summation over
repeated indices implied)
\bea
\Gamma^L_{ll'} & =& B^{L}_{li}B^{L\ast}_{l'i} \Bigg\{\delta_{ij}
\Bigg[c^2_w\lZ\Big(C_{11}(\li,1,1)+C_{23}(\li,1,1)-C_{22}(\li,1,1)\Big)
\nonumber\\
&&+ 6c^2_wC_{24}(\li,1,1)-s^2_w\li C_0(\li,1,1) + \frac{1}{2}(1-2s^2_w)
\Big(\li C_{24}(\li,1,1) \nonumber\\
&&+ \frac{1}{2}\li B_1(0,\li,1) + B_1(0,\li,1)\Big)\Bigg]\nonumber\\
&&+ C^{L}_{ij}\Bigg[ -C_{24}(1,\li,\lj)-\frac{1}{2}\lZ\Big( C_0(1,\li,\lj)
+ C_{11}(1,\li,\lj)\nonumber\\
&&+ C_{23}(1,\li,\lj)-C_{22}(1,\li,\lj)\Big)
- \frac{1}{4}\li\lj C_0(1,\li,\lj)\Bigg]\nonumber\\
&&+ \frac{1}{2}C^{L\ast}_{ij}\sqrt{\li\lj}\Bigg[ C_0(1,\li,\lj)+
\frac{1}{2}\lZ \Big(C_{23}(1,\li,\lj)-C_{22}(1,\li,\lj)\Big)+
C_{24}(1,\li,\lj)\Bigg]\nonumber\\
&& +\frac{1}{4}C^{R}_{ij}\sqrt{\li\lj}\Bigg[2C_{24}(0,\lS,\lI)-
C_{24}(\lS,0,0)-C_{24}(\lI,0,0)+\frac{1}{2}\nonumber\\
&&+ s^2_w\Big( B_1(0,0,\lS) + B_1(0,0,\lI)\Big) -\, \frac{1}{2}\lZ\Big(
C_{23}(\lS,0,0)-C_{22}(\lS,0,0) \nonumber\\
&&+ C_{23}(\lI,0,0)-C_{22}(\lI,0,0)\Big) \Bigg]\, \Bigg\}\, ,
\eea
where $\li=m^2_i/M^2_W$, $\lZ=M^2_Z/M^2_W$, $\lS=M^2_{\phi^{0r}_2}/M^2_W$,
and $\lI=M^2_{\phi^{0i}_2}/M^2_W$. Note that there is a contribution
proportional to $C^R_{ij}$ that originates solely from the Higgs sector
of the LRSM. In the notation of~\cite{Bernd}, the first three of the
six arguments of the $C$ functions are always evaluated at $(0,\lZ,0)$.

In the LRSM, virtual neutrinos and Higgs scalars induce a nonuniversal $Z$
boson coupling to right-handed charged leptons, $\Gamma^R$, as
shown in Fig.~1. The contributions of the individual graphs  to $\Gamma^R$
are listed below
\bea
\Gamma^R_{ll'}(a) &=& B^R_{li}B^{R\ast}_{l'i} s^2_w\Bigg[\lZ\Big(
C_{22}(\li,\lR,\lR) - C_{23}(\li,\lR,\lR) -C_{11}(\li,\lR,\lR) \Big)\nonumber\\
&&-6C_{24}(\li,\lR,\lR)\Bigg],\\
\Gamma^R_{ll'}(b+c) &=& B^R_{li}B^{R\ast}_{l'i} (s^2_w-s^2_\beta)\li
C_0(\li,\lR,\lR),\\
\Gamma^R_{ll'}(d) &=& \frac{1}{2}B^R_{li}B^{R\ast}_{l'i} s^2_\beta
(s^2_\beta-2s^2_w)\li C_{24}(\li,\lR,\lR),\\
\Gamma^R_{ll'}(e+f) &=& -\frac{1}{2}B^R_{li}B^{R\ast}_{l'j}
(\delta_{ij}s^2_\beta\li -\sqrt{\li\lj} C^L_{ij})\Big( C_0(\li,\lR,\lh)+
C_0(\lj,\lR,\lh)\Big),\\
\Gamma^R_{ll'}(g) &=& \frac{1}{2}B^R_{li}B^{R\ast}_{l'j}\sqrt{\lk\ln}\left(
1-\frac{2s^2_w}{c^2_\beta}\right) (s^2_\beta\delta_{ki} -
C^L_{ki})(s^2_\beta\delta_{in}-C^L_{in})C_{24}(\li,\lh,\lh),\\
\Gamma^R_{ll'}(h+i) &=& -\frac{1}{2}B^R_{li}B^{R\ast}_{l'j}\sqrt{\li\lj}
s^2_\beta (s^2_\beta\delta_{ij} -C^L_{ij})\Big( C_{24}(\li,\lR,\lh)+
C_{24}(\lj,\lR,\lh)\Big),\\
\Gamma^R_{ll'}(j) &=& -\frac{1}{2}B^R_{li}B^{R\ast}_{l'j}\Bigg\{
C^{L\ast}_{ij}\Bigg[ 1 - 2 C_{24}(\lR,\li,\lj) -\lZ\Big(
C_0(\lR,\li,\lj) + C_{11}(\lR,\li,\lj)\nonumber\\
&& +C_{23}(\lR,\li,\lj) - C_{22}(\lR,\li,\lj) \Big) \Bigg]\ +
C^L_{ij}\sqrt{\li\lj}C_0(\lR,\li,\lj) \Bigg\},\\
\Gamma^R_{ll'}(k) &=& -\frac{1}{4}B^R_{li}B^{R\ast}_{l'j}s^2_\beta
\sqrt{\li\lj} \Bigg\{ C^L_{ij}\Bigg[ 2C_{24}(\lR,\li,\lj) -\frac{1}{2}+\lZ\Big(
C_{23}(\lR,\li,\lj)\nonumber\\
&&- C_{22}(\lR,\li,\lj) \Big) \Bigg]
- C^{L\ast}_{ij}\sqrt{\li\lj}C_0(\lR,\li,\lj) \Bigg\},\\
\Gamma^R_{ll'}(l) &=& -\frac{1}{4}B^R_{lk}B^{R\ast}_{l'n}
\sqrt{\lk\ln} \frac{1}{c^2_\beta}(s^2_\beta\delta_{ki}-C^L_{ki})
(s^2_\beta\delta_{nj}- C^L_{jn} )
\Bigg\{ C^L_{ij}\Bigg[ 2C_{24}(\lh,\li,\lj) -\frac{1}{2}\nonumber\\
&&+\lZ\Big( C_{23}(\lh,\li,\lj) - C_{22}(\lh,\li,\lj) \Big) \Bigg]
- C^{L\ast}_{ij}\sqrt{\li\lj}C_0(\lh,\li,\lj) \Bigg\}\ \\
\Gamma^R_{ll'}(m+n) &=& -\frac{1}{2}B^R_{li}B^{R\ast}_{l'j}
C^L_{ij}\sqrt{\li\lj}C_{24}(0,\lS,\lI),\\
\Gamma^R_{ll'}(o+p) &=& \frac{1}{8}B^R_{li}B^{R\ast}_{l'j}
C^L_{ij}(1-2s^2_w)\sqrt{\li\lj} \Bigg[ 2C_{24}(\lS,0,0)+ 2C_{24}(\lI,0,0)
         -1\nonumber\\
&& + \lZ\Big( C_{23}(\lS,0,0)-C_{22}(\lS,0,0)+
C_{23}(\lI,0,0)-C_{22}(\lI,0,0) \Big) \Bigg], \\
\Gamma^R_{ll'}(q) &=& \frac{1}{2}B^R_{li}B^{R\ast}_{l'j}
C^{R\ast}_{ij}s^2_w\frac{s^2_\beta}{c^2_\beta}
\sqrt{\li\lj}C_{24}(0,\ld,\ld),\\
\Gamma^R_{ll'}(r) &=& -\frac{1}{8}B^R_{li}B^{R\ast}_{l'j}
C^{R\ast}_{ij}s^2_w\frac{s^2_\beta}{c^2_\beta}
\sqrt{\li\lj} \Bigg[ 2C_{24}(\ld,0,0)- \frac{1}{2}+\lZ\Big(
C_{23}(\ld,0,0)\nonumber\\
&&-C_{22}(\ld,0,0) \Big) \Bigg],
\eea
where $\lambda_h=M^2_{h^+}/M^2_W$ and $\ld=M^2_{\delta^{++}_R}/M^2_W$.
In addition to the irreducible three-point functions, we should take
wave-function renormalization constants into account (Figs.~1(A)--(F)).
These additional nonuniversal corrections generated by the selfenergies
are calculated to give
\bea
\Gamma^R_{ll'}(A) &=& -\frac{1}{2}B^R_{li}B^{R\ast}_{l'j}
s^2_w \Big(1+2B_1(0,\li,\lR)\Big), \\
\Gamma^R_{ll'}(B) &=& -\frac{1}{2}B^R_{li}B^{R\ast}_{l'j}
s^2_w s^2_\beta\li B_1(0,\li,\lR), \\
\Gamma^R_{ll'}(C) &=& -\frac{1}{2}B^R_{li}B^{R\ast}_{l'j}
\frac{s^2_w}{c^2_\beta} \sqrt{\lk\ln} (s^2_\beta\delta_{ki} -C^L_{ki})
(s^2_\beta\delta_{in} -C^L_{in})B_1(0,\li,\lh),\\
\Gamma^R_{ll'}(D+E) &=& -\frac{1}{4}B^R_{li}B^{R\ast}_{l'j}C^L_{ij}
s^2_w \sqrt{\li\lj} \Big( B_1(0,0,\lS) + B_1(0,0,\lI) \Big),\\
\Gamma^R_{ll'}(F) &=& \frac{1}{8}B^R_{li}B^{R\ast}_{l'j}C^{R\ast}_{ij}
s^2_w\frac{s^2_\beta}{c^2_\beta} \sqrt{\li\lj} B_1(0,0,\ld).
\eea
The sum of Eqs.~(B.2)--(B.19) should be free from UV divergences,
when $l\neq l'$. This can easily be verified by employing the
identities that the mixing matrices $B^{L,R}$ and $C^{L,R}$
obey (see also discussion in Section 2). An ultimate check for
the correctness of our analytic results is the vanishing of
all terms involving $s^2_w$ in the limit $\lZ\to 0$, due to
electromagnetic gauge invariance.

\end{appendix}
\newpage


\newpage

\centerline{\bf\Large Figure Captions }
\vspace{-0.2cm}
\newcounter{fig}
\begin{list}{\bf\rm Fig. \arabic{fig}: }{\usecounter{fig}
\labelwidth1.6cm \leftmargin2.5cm \labelsep0.4cm \itemsep0ex plus0.2ex }

\item Feynman graphs contributing to the effective nonoblique
$Z l_R\bar{l}_R$ coupling in the LRSM.

\item $B(Z \to l_1\bar{l}_2 + \bar{l}_1l_2, l_1\not= l_2)$ in a
two-generation LRSM
as a function of {\bf (a)} the heavy neutrino mass $m_N(=m_{N_1}=m_{N_2})$,
{\bf (b)} $W_R$-boson mass $M_R$, {\bf (c)} charged Higgs boson $M_h$,
[for the $L$-violating mixings  $(s^{\nu_\tau}_L)^2=0.04$~(curve-$i$),
0.03~(curve-$ii$), 0.020~(curve-$iii$) and setting $(s^{\nu_e}_L)^2=0.01$,
$(s^{\nu_\mu}_L)^2=0$], and
{\bf (d)} a Cabbibo-type angle $\theta_R$ [for $M_R=0.4$~TeV (curve-$i$),
0.6~TeV (curve-$ii$), and 0.8~TeV (curve-$iii$)]. Numerical estimates coming
solely from the $SU(2)_R$ sector are also shown. The results analogous
to the curves--$i$, $ii$, and~$iii$, are correspondingly given by the
dashed, dotted, and dash-dotted lines.

\item Numerical estimates of $U_{br}^{l_1l_2}$ for the same set
of parameters as in Fig.~2.

\item Numerical estimates of $\Delta\cA_{l_1l_2}$ for the same
set of parameters as in Fig.~2. Only the total LRSM contribution
to $\Delta\cA_{l_1l_2}$ is shown, where
the corresponding curves--$i$, $ii$, and~$iii$ are now given by the
solid, dashed, and dotted lines, respectively.

\end{list}

\end{document}